\documentclass{article}
\usepackage{emulateapj,pstricks,apjfonts}
\def\asca       {{\em ASCA}\/}
\def\chandra    {{\em Chandra}\/}
\def\ginga    	{{\em Ginga}\/}
\def\xmm        {{\em XMM}\/}
\def\rosat      {{\em ROSAT}\/}
\def\am         {$^\prime$}
\def\as         {$^{\prime\prime}$}
\def\cmsq       {~cm$^{-2}$}
\def\kms        {~km$\;$s$^{-1}$}
\def\kmsmpc     {~km$\;$s$^{-1}\,$Mpc$^{-1}$}
\def\msunyr     {$M_{\odot}\;$yr$^{-1}$}
\def\degd       {$^{\circ}\!$}
\begin{document}


\lefthead{CHANDRA OBSERVATION OF ABELL 2142}
\righthead{MARKEVITCH ET AL.}

\title{{\em CHANDRA}\/ OBSERVATION OF ABELL 2142: SURVIVAL OF DENSE
SUBCLUSTER CORES IN A MERGER}

\author{M. Markevitch$^1$, T.~J.~Ponman$^{1,2}$, P.~E.~J.~Nulsen$^{1,3}$,
M.~W.~Bautz$^4$, D.~J.~Burke$^5$, L.~P.~David$^1$, D.~Davis$^4$,
R.~H.~Donnelly$^1$, W.~R.~Forman$^1$, C.~Jones$^1$, J.~Kaastra$^6$,
E.~Kellogg$^1$, D.-W.~Kim$^1$, J.~Kolodziejczak$^7$, P.~Mazzotta$^{1,8}$,
A.~Pagliaro$^5$, S.~Patel$^6$, L.~Van~Speybroeck$^1$, A.~Vikhlinin$^{1,9}$,
J.~Vrtilek$^1$, M.~Wise$^4$, P.~Zhao$^1$}

\setcounter{footnote}{9}

\footnotetext[1]{Harvard-Smithsonian Center for Astrophysics, 60 Garden St.,
Cambridge, MA 02138
~$^2$University of Birmingham, U.K.
~$^3$University of Wollongong, Australia
~$^4$MIT
~$^5$IfA, University of Hawaii
~$^6$SRON, the Netherlands
~$^7$NASA/MSFC
~$^8$Universita di Roma, Italy
~$^9$IKI, Russian Academy of Sciences}

\begin{abstract}

We use \chandra\ data to map the gas temperature in the central region of
the merging cluster A2142. The cluster is markedly nonisothermal; it appears
that the central cooling flow has been disturbed but not destroyed by a
merger. The X-ray image exhibits two sharp, bow-shaped, shock-like surface
brightness edges or gas density discontinuities. However, temperature and
pressure profiles across these edges indicate that these are not shock
fronts. The pressure is reasonably continuous across these edges, while the
entropy jumps in the opposite sense to that in a shock (i.e.\ the denser
side of the edge has lower temperature, and hence lower entropy).  Most
plausibly, these edges delineate the dense subcluster cores that have
survived a merger and ram pressure stripping by the surrounding shock-heated
gas.

\end{abstract}


\section{INTRODUCTION}

Clusters of galaxies grow through gravitational infall and merger of smaller
groups and clusters. During a merger, a significant fraction of the enormous
($\sim 10^{63-64}$ ergs) kinetic energy of the colliding subclusters
dissipates in the intracluster gas through shock heating, giving rise to
strong, but transient, spatial variations of gas temperature and entropy.
These variations contain information on the stage, geometry and velocity of
the merger. They also can shed light on physical processes and phenomena
occurring in the intracluster medium, including gas bulk flows, destruction
of cooling flows, turbulence, and thermal conduction. Given this wealth of
information contained in the merger temperature maps, they have in the past
few years been a subject of intensive study, both experimental (using
\rosat\ PSPC and \asca\ data, e.g., Henry \& Briel 1996; Markevitch,
Sarazin, \& Vikhlinin 1999, and references in those works) and theoretical,
using hydrodynamic simulations (e.g., Schindler \& Muller 1993; Roettiger,
Burns, \& Stone 1999 and references therein). The measurements reported so
far, while revealing, were limited by the \rosat's limited energy coverage
and the \asca's moderate angular resolution. Two new X-ray observatories,
\chandra\ and \xmm, will overcome these difficulties and provide much more
accurate spatially resolved temperature data, adequate for studying the
above phenomena.

In this paper, we analyze the first \chandra\ observation of a merging
cluster, A2142 ($z=0.089$). This hot ($T_e\sim 9$ keV), X-ray-luminous
cluster has two bright elliptical galaxies near the center, aligned in the
general direction of the X-ray brightness elongation. The line-of-sight
velocities of these galaxies differ by $1840$\kms\ (Oegerle, Hill, \&
Fitchett 1995), suggesting that the cluster is not in a dynamically relaxed
state. The X-ray image of the cluster has a peak indicating a cooling flow.
From the \rosat\ HRI image, Peres et al.\ (1998) deduced a cooling flow rate
of $72^{+14}_{-19}\,h^{-2}$\msunyr.  From the \rosat\ PSPC image, Buote \& Tsai
(1996) argued that this cluster is at a late merger stage.  Henry \& Briel
(1996) used \rosat\ PSPC data to derive a rough gas temperature map for
A2142.  Since this cluster is too hot for the PSPC to derive accurate
temperatures, they adjusted the PSPC gain to make the average temperature
equal to that from \ginga\ and looked for spatial hardness variations. Their
temperature map showed azimuthally asymmetric temperature variations, which
also is an indication of a merger.  A derivation of an
\asca\ temperature map for this relatively distant cluster was hindered by
the presence of a central brightness peak associated with a cooling flow.

Examination of the \rosat\ PSPC and HRI images reveals two striking X-ray
brightness edges within a few arcminutes northwest and south of the
brightness peak, which were not reported in the earlier studies of A2142.
The new \chandra\ data show these intriguing cluster gas features more
clearly and allow us to study them in detail, including spectroscopically.
\chandra\ also provides a high-resolution temperature map of the central
cluster region. These results are presented below. We use $H_0=100\,h$
\kmsmpc\ and $q_0=0.5$; confidence intervals are one-parameter 90\%.

\section{DATA REDUCTION}
\label{sec:data}

A2142 was observed by \chandra\ during the calibration phase on 1999 August
20 with the ACIS-S detector
\footnote{\chandra\ Observatory Guide
http://asc.harvard.edu/udocs/docs/docs.html,
section ``Observatory Guide'', ``ACIS''}.
Two similar, consecutive observations
(OBSID 1196 and 1228) are combined here.  The data were telemetered in Faint
mode.  Known hot pixels, bad columns, chip node boundaries, and events with
\asca\ grades 1, 5, and 7 are excluded from the analysis, along with several
short time intervals with incorrect aspect reconstruction. The cluster was
centered in the backside-illuminated chip S3 that is susceptible to particle
background flares%
\footnote{\chandra\ memo
http://asc.harvard.edu/cal/Links/Acis/acis/WWWacis\_cal.html, 
section ``Particle Background''}.
For our study of the low surface brightness regions of the cluster, it is
critical to exclude any periods with anomalous background. For that, we made
a light curve for a region covering 1/5 of the S3 chip far from the cluster
peak where the relative background contribution to the flux is largest,
using screened events in the 0.3--10 keV energy band (Fig.\ 1). The light
curve shows that most of the time the background is quiescent (approximately
half of the flux during these periods is due to the cluster emission in this
region of the detector) but there are several flares. We excluded all time
intervals when the flux was significantly, by more than $3\sigma$, above or
below the quiescent rate (the flux may be below normal, for example, due to
data dropouts).  The excluded intervals are shaded in Fig.\ 1. This
screening resulted in a total clean exposure of 16.4 ks for the S3 chip (out
of a total of 24 ks).  The same flare intervals can be identified from the
light curve of another backside-illuminated chip, S1, that also was active
during the exposure but has a much smaller cluster contribution. A similar
screening of the frontside-illuminated chips, less affected by the flares,
resulted in a total clean exposure of 21.3 ks for those chips.  In this
paper, we limit our imaging analysis to chips S2 and S3 and spectral
analysis to chip S3.

During the quiescent periods, the particle background is rather constant in
time but is non-uniform over the chip (varying by $\sim 30$\% on scales of a
few arcmin). To take this nonuniformity into account in our spectral and
imaging analysis, we used a background dataset composed of several other
observations of relatively empty fields with bright sources removed. Those
observations were screened in exactly the same manner as the cluster data.
The total exposure of that dataset is about 70 ks. To be able to extract the
background spectra and images in sky coordinates corrected for the
observatory dither, chip coordinates of the events from the background
dataset were converted to the sky coordinate frame of the observation being
analyzed.  This was done by assigning randomly generated time tags to the
background events and applying the corresponding aspect correction. The
background spectra or images were then normalized by the ratio of the
respective exposures.  This procedure yields a background which is accurate
to $\sim 10$\% based on comparison to other fields; this uncertainty will be
taken into account in our results.

For generating a temperature map (\S\ref{sec:tmap}), we corrected the images
for the effect of the source smearing during the periods of CCD frame
transfer. While the frame transfer duration (41 ms) is small compared to the
useful exposure (3.2 s) in each read-out cycle, the contamination may be
significant for the outer, low surface brightness regions of the cluster
that have the same chip $x$ coordinates as the cluster sharp brightness
peak. To a first approximation, this effect can be corrected by convolving
the ACIS image with the readout trajectory (a line parallel to the chip $y$
axis), multiplying by the ratio of the frame transfer and useful exposures,
and subtracting from the uncorrected image. This assumes that the image is
not affected by the pileup effect, which is true for most cluster data,
including ours.

\section{RESULTS}

\subsection{Image}

An ACIS image of the cluster using the 0.3-10 keV events from chips S2 and
S3 is shown in Fig.\ 2 (the cluster peak is in S3).  An overlay of the X-ray
contours on the DSS optical plate in Fig.\ 2{\em b}\/ shows that the cluster
brightness peak is slightly offset from the central galaxy (galaxy 201 in
the Oegerle et al.\ notation; we will call it G1), and that the second
bright galaxy, hereafter G2 (or galaxy 219 from Oegerle et al.), does not
have any comparable gas halo around it. North of G2, there is an X-ray point
source coincident with a narrow-tail radio galaxy (Harris, Bahcall, \& Strom
1977).

The image in Fig.\ 2{\em a}\/ shows a very regular, elliptical brightness
distribution and two striking, elliptical-shaped edges, or abrupt drops, in
the surface brightness, one $\sim 3'$ northwest of the cluster center and
another $\sim 1'$ south of the center. We derive gas density and
temperature profiles across these interesting structures in
\S\S\ref{sec:tprof}-\ref{sec:dprof}.

\subsection{Average cluster spectrum}
\label{sec:avg}

Before proceeding to the spatially-resolved spectroscopy, we fit the overall
cluster spectrum to check the consistency with previous studies. For this,
we use a spectrum from the entire S3 chip, excluding point sources. This
approximately corresponds to an integration radius of 5\am. At present, the
soft spectral response of the S3 chip is uncertain and we observe
significant residual deviations below $E\simeq 0.7$ keV for any reasonable
spectral models. Therefore, we have chosen to restrict all spectral analysis
to energies 1--10 keV. The cluster is hot and this choice does not limit the
accuracy of our main results. The spectra were extracted in PI (pulse
height-invariant) channels that correct for the gain difference between the
different regions of the CCD. The spectra from both pointings were grouped
to have a minimum of 100 counts per bin and fitted simultaneously using the
{\small XSPEC} package (Arnaud 1996). Model spectra were multiplied by the
vignetting factor (auxiliary response) calculated by weighting the
position-dependent effective area with the X-ray brightness over the
corresponding image region. Fitting results for an absorbed
single-temperature thin plasma model (Raymond \& Smith 1977, 1992 revision)
and a model with an additional cooling flow component are given in Table 1,
where the iron abundance is relative to that of Anders \& Grevesse (1989).
Our single-temperature fit is in reasonable agreement with values from
\ginga\ ($9.0\pm0.3$ keV for $N_H=5\times10^{20}$\cmsq; White et al.\ 1994)
and \asca\ ($8.8\pm0.6$ keV for $N_H=4.2\times10^{20}$\cmsq; Markevitch et
al.\ 1998). At this stage of the \chandra\ calibration, and for our
qualitative study, the apparent small discrepancy is not a matter of
concern; also, the above values correspond to different integration regions
for this highly non-isothermal cluster. If we allow for a cooling flow
component (see Table 1), our temperature is consistent with a similarly
derived \asca\ value, $9.3^{+1.3}_{-0.7}$ keV (Allen \& Fabian 1998), and
the cooling rate with the one derived from the \rosat\ images (Peres et al.\
1998), although the presence of a cooling flow is not strongly required by
the overall spectrum in our energy band. The table also shows that the
absorbing column is weakly constrained (due to our energy cut) but is in
good agreement with the Galactic value of $4.2\times 10^{20}$\cmsq (Dickey
\& Lockman 1990). We therefore fix $N_H$ at its Galactic value in the
analysis below.

{\footnotesize
\renewcommand{\arraystretch}{1.4}
\renewcommand{\tabcolsep}{1mm}
\begin{center}
TABLE 1
\vspace{1mm}

{\sc Overall Spectrum Fits}
\vspace{1mm}

\begin{tabular}{cccccc}
\hline \hline
Model       & $T_e$, & $N_H$,  & Abund. & $\dot{M}$, & $\chi^2$/d.o.f \\
	    & keV    & $10^{20}$\cmsq   & & $h^{-2}\,$\msunyr&       \\
\hline
single-$T$  & $8.1\pm0.4$         & $3.8\pm1.5$  & $0.27\pm0.04$ 
  & ...                 & 517.2 / 493 \\
cooling flow& $8.8^{+1.2}_{-0.9}$ & $5.9\pm2.8$  & $0.28\pm0.04$ 
  & $69^{+70}_{-...}$ & 515.1 / 492 \\
\hline
\end{tabular}
\end{center}
}
\vspace{3mm}

\subsection{Temperature map}
\label{sec:tmap}

Using \chandra\ data, it is possible to derive a two-dimensional temperature
map within $3'-4'$ of the cluster peak. The \chandra\ angular resolution is
more than sufficient to allow us to ignore any energy-dependent PSF effects
and, for example, simply convert an X-ray hardness ratio at each cluster
position to temperature. Taking advantage of this simplicity, we also tried
to use as much spectral information as possible without dividing the cluster
into any regions for full spectral fitting. To do this, we extracted images
in five energy (or PI) bands 1.0--1.5--2.0--3.0--5.5--10 keV, smoothed them,
and for each $2''\times 2''$ pixel fitted a spectrum consisting of the flux
values in each band properly weighted by their statistical errors. The
corresponding background images were created as described in
\S\ref{sec:data} and subtracted from each image. The background-subtracted
images were approximately corrected for the frame transfer smearing effect
following the description in \S\ref{sec:data} and divided by the vignetting
factor relative to the on-axis position (within each energy band, the
vignetting factor for different energies was weighted using a 10 keV plasma
spectrum). The images were then smoothed by a variable-width Gaussian (same
for all bands) whose $\sigma$ varied from 10\as\ at the cluster peak to
30\as\ near the edges of the map. Bright point sources were masked prior to
smoothing. We used a one-temperature plasma model with the absorption column
fixed at the Galactic value and iron abundance at the cluster average,
multiplying the model by the on-axis values of the telescope effective area
(since the images were vignetting-corrected). The instrument spectral
response matrix was properly binned for our chosen energy bands.

The resulting temperature map is shown in Fig.\ 3. The useful exposure of
our observations is relatively short so the statistical accuracy is limited.
The map shows that the cluster brightness peak is cool and that this cool
dense gas is displaced to the SE from the main galaxy G1. There is also a
cool filament extending from the peak in the general direction of the second
galaxy G2, or along the southern brightness edge. The G2 galaxy itself is
not associated with any features in the temperature map.

At larger scales, the map shows that the hottest cluster gas lies
immediately outside the NW brightness edge and to the south of the southern
edge. In the relatively small region of the cluster covered by our analysis,
our temperature map is in general agreement with the coarser \rosat/\ginga\
map of Henry \& Briel (1996). Both maps show that the center of the cluster
is cool (probably has a cooling flow) and the hot gas lies outside, mostly
to the north and west. The maps differ in details; for example, our map
indicates an increase of the temperature southeast of the center where the
\rosat\ map suggests a decrease. An important conclusion from our map is
that the brightness edges separate regions of cool and hot gas. These edges
are studied in more detail in sections below.

There is also some marginal evidence in Fig.\ 3 for a faint cool filament
running across the whole map through the cluster brightness peak and
coincident with the chip quadrant boundary. It is within the statistical
uncertainties and most probably results from some presently unknown detector
effect. This feature does not affect our arguments.

\subsection{Temperature profiles across the edges}
\label{sec:tprof}

To derive the temperature profiles across the edges, we divide the cluster
into elliptical sectors as shown in Fig.\ 4{\em a}, chosen so that the
cluster edges lie exactly at the boundaries of certain sectors, and so that
the sectors cover the azimuthal angles where the edges are most prominent.
Figure 4{\em b} shows the best-fit temperature values in each region, for
both observations fitted together or separately (for a consistency check).
The fitting was performed as described in \S\ref{sec:avg}. The temperatures
shown in the figure correspond to the iron abundance fixed at the cluster's
average and a fixed Galactic absorption; when fit as a free parameter, the
absorption column was consistent with the Galactic value in all regions. For
both edges, as we move from the inside of the edge to the outer, less dense
region, the temperature increases abruptly and significantly. The profiles
also show a decrease of the temperature in the very center of the cluster,
which is also seen in the temperature map in Fig.\ 3.

We must note here that our spectral results in the outer, low surface
brightness regions of the cluster depend significantly on the background
subtraction. To quantify the corresponding uncertainty, we varied the
background normalization by $\pm10$\% (synchronously for the two
observations), re-fitted the temperatures in all sectors and added the
resulting difference in quadrature to the 90\% statistical uncertainties.
While the values for the brighter cluster regions are practically
unaffected, for the regions on the outer side of the NW edge, these
differences are comparable to the statistical uncertainty. The 10\% estimate
is rather arbitrary and appears to overestimate the observed variation of
the ACIS quiescent particle background with time. A possible incomplete
screening of background flares is another source of uncertainty that is
difficult to quantify.  Experimenting with different screening criteria
shows that it can significantly affect the results. An approximate estimate
of this uncertainty is made by comparing separate fits to the two
observations (dotted crosses in Fig.\ 4{\em b}); their mutual consistency
shows that for the conservative data screening that we used, this
uncertainty is probably not greater than the already included error
components.

\subsection{Density and pressure profiles}
\label{sec:dprof}

Figure 4{\em c} shows X-ray surface brightness profiles across the two
edges, derived using narrow elliptical sectors parallel to those used above
for the temperatures. The energy band for these profiles is restricted to
0.5--3 keV to minimize the dependence of X-ray emissivity on temperature and
to maximize the signal-to-noise ratio. Both profiles clearly show the sharp
edges; the radial derivative of the surface brightness is discontinuous on a
scale smaller than $5''-10''$ (or about $5-10\,h^{-1}$ kpc, limited mostly
by the accuracy with which our regions can be made parallel to the edges).
The brightness edges have a very characteristic shape that indicates a
discontinuity in the gas density profile. To quantify these discontinuities,
we fitted the brightness profiles with a simple radial density model with
two power laws separated by a jump. The curvature of the edge surfaces along
the line of sight is unknown; therefore, for simplicity, we projected the
density model under the assumption of spherical symmetry with the average
radius as the single radial coordinate, even though the profiles are derived
in elliptical regions. The accuracy of such modeling is sufficient for our
purposes. We also restrict the fitting range to the immediate vicinity of
the brightness edges (see Fig.\ 4{\em c}) and ignore the gas temperature
variations since they are unimportant for the energy band we use. The free
parameters are the two power-law slopes and the position and amplitude of
the density jump. The best-fit density models are shown in Fig.\ 4{\em d}
and the corresponding brightness profiles are overlaid as histograms on the
data points in Fig.\ 4{\em c}. The best-fit amplitudes of the density jumps
are given by factors of $1.85\pm0.10$ and $2.0\pm0.1$ for the S and NW
edges, respectively. As Figure 4{\em c} shows, the fits are very good, with
respective $\chi^2=26.5/25$ d.o.f.\ and $18.9/22$ d.o.f. The goodness of
fits suggests that the curvature of the edges along the line of sight is
indeed fairly close to that in the plane of the sky. To estimate how
model-dependent are the derived amplitudes of the jumps, we tried to add a
constant density background (positive or negative) as another fitting
component representing possible deviations of the profile from the power law
at large radii. The resulting changes of the best-fit jump amplitudes were
comparable to the above small uncertainties.  Thus, our evaluation of the
density discontinuities appears robust, barring strong projection effects
that can reduce the apparent density jump at the edge.

From the density and temperature distributions in the vicinity of the
brightness edges, we can calculate the pressure profiles. Note that even
though the measured temperatures correspond to emission-weighted projections
along the line of sight, they are reasonably close to the true
three-dimensional temperatures at any given radius, because the X-ray
brightness declines steeply with radius. Figure 4{\em e} shows pressure
profiles calculated by multiplying the measured temperature values and the
model density values in each region (the density is taken at the
emission-weighted radius for each region). Remarkably, while the temperature
and density profiles both exhibit clear discontinuities at the edges, the
pressure profiles are consistent with no discontinuity within the
uncertainties. Thus the gas is close to local pressure equilibrium at the
density edges. It is also noteworthy that the denser gas inside the edges
has lower specific entropy, therefore the edges are convectively stable.

\section{DISCUSSION}

Shock fronts would seem the most natural interpretation for the density
discontinuities seen in the X-ray image of A2142. Such an interpretation was
proposed for a similar brightness edge seen in the \rosat\ image of another
merging cluster, A3667 (Markevitch, Sarazin, \& Vikhlinin 1999), even though
the \asca\ temperature map did not entirely support this explanation.
However, if these edges in A2142 were shocks, they would be accompanied by a
temperature change across the edge in the direction opposite to that
observed. Indeed, applying the Rankine--Hugoniot shock jump conditions for a
factor of $\sim 2$ density jump and taking the post-shock temperature to be
$\sim 7.5$ keV (the inner regions of the NW edge), one would expect to find
a $T\simeq 4$ keV gas in front of the shock (i.e. on the side of the edge
away from the cluster center). This is inconsistent with the observed clear
increase of the temperature across both edges and the equivalent increase of
the specific entropy. This appears to exclude the shock interpretation. An
alternative is proposed below.

\subsection{Stripping of cool cores by shocked gas}
\label{sec:d1}

The smooth, comet-like shape and sharpness of the edges alone (especially of
the NW edge) may hint that we are observing a body of dense gas moving
through and being stripped by a less dense surrounding gas. This dense body
may be the surviving core of one of the merged subclusters that has not been
penetrated by the merger shocks due to its high initial pressure.  The edge
observed in the X-ray image could then be the surface where the pressure in
the dense core gas is in balance with the thermal plus ram pressure of the
surrounding gas; all core gas at higher radii that initially had a lower
pressure has been stripped and left behind (possibly creating a tail seen as
a general elongation to the SE). The hotter, rarefied gas beyond the NW edge
can be the result of shock heating of the outer atmospheres of the two
colliding subclusters, as schematically shown in Fig.\ 5. In this scenario,
the outer subcluster gas has been stopped by the collision shock, while the
dense cores (or, more precisely, regions of the subclusters where the
pressure exceeded that of the shocked gas in front of them, which prevented
the shock from penetrating them) continued to move ahead through the shocked
gas.  The southern edge may delineate the remnant of the second core (core B
in Fig.\ 5) that was more dense and compact and still retains a cooling
flow.  The two cores should have already passed the point of minimum
separation and be moving apart at present. It is unlikely that the less
dense core A could survive a head-on passage of the denser core (in that
case we probably would not see the NW edge). This suggests a nonzero impact
parameter; for example, the cores could have been separated along the line
of sight during the passage, with core B either grazing or being projected
onto core A at present.

Although the thermal pressure profiles in Fig.\ 4{\em e} do not suggest any
abrupt decline across the edges that could be due to a ram pressure
component, at the present accuracy they do not strongly exclude it. To
estimate what bulk velocity, $\upsilon$, is consistent with the data on the
NW edge, we can apply the pressure equilibrium condition to the edge
surface, $p_1=p_2+\rho_2 \upsilon^2$, where indices 1 and 2 correspond to
quantities inside and outside the edge, respectively. The density jump by a
factor of 2 and the 90\% lower limit on the temperature in the nearest outer
bin, $T_2>10$ keV, corresponds to an average bulk velocity of the gas in
that region of $\upsilon<900$\kms. This is consistent with subcluster
velocities of order 1000\kms\ expected in a merger such as A2142. Note,
however, that this is a very rough estimate because, if our interpretation
is correct, the gas velocity would be continuous across the edge and there
must be a velocity gradient, as well as a compression with a corresponding
temperature increase, immediately outside the edge. Also, if the core moves
at an angle to the plane of the sky, the maximum velocity may be higher,
since one of its components would be tangential to the contact surface that
we can see.  In addition, as noted above, projection effects can dilute the
density jump, leaving smaller apparent room for ram pressure. A similar
estimate for the southern edge is $\upsilon<400$\kms, but it is probably
even less firm because of the likely projection of core B onto core A.

Depending on the velocities of the cores relative to the surrounding
(previously shocked) gas, they may or may not create additional bow shocks
at some distance in front of the edges (shown as dashed lines in Fig.\ 5).
The above upper limit on the velocity of core A is lower than the sound
velocity in a $T>10$ keV gas ($\upsilon_s>1600$\kms) and is therefore
consistent with no shock, although it does not exclude it due to the
possible projection and orientation effects mentioned above. The available
X-ray image and temperature map do not show any obvious corresponding
features, but deeper exposures might reveal such shocks.

Comparison of the X-ray and optical images (Fig. 2{\em b}) offers an
attractive possibility that core B is centered on galaxy G1 and core A on
galaxy G2. However, this scenario has certain problems. Velocity data, while
scarce ($\sim 10$ galaxy velocities in the central region; Oegerle et al.),
show that G2 is separated from most other cluster members by a line-of-sight
velocity of $\sim 1800$\kms, except for the radio galaxy north of G2 that
has a similar velocity. It therefore appears unlikely that G2 can be the
center of a relatively big, $T\simeq 7$ keV subcluster, unless a deeper
spectroscopic study reveals a concentration of nearby galaxies with similar
velocity. It is possible that this galaxy is completely unrelated to core A;
we recall that it does not display any strong X-ray brightness enhancement.
Another problem is a displacement, in the wrong direction, of the cool
density peak from the G1 galaxy. If G1 is at the peak of the gravitational
potential of the smaller core B, one would expect the gas to lag behind the
galaxy as the core moves to the south or southeast (as the edge suggests).
The observed displacement might be explained if at present core B is moving
mostly along the line of sight on a circular orbit and the central galaxy is
already starting its turnaround toward G2, perhaps leaving behind a trail of
cool gas seen as a cool filament. The observed southern edge would then be a
surface where the relative gas motion is mostly tangential, which is also in
better agreement with the low allowed ram pressure.

Below we propose a slightly different scenario for the merger, motivated by
comparison of the observed structure with hydrodynamic simulations. It
invokes the same physical mechanism for the observed density edges.

\subsection{Late stage unequal mass merger}
\label{sec:d2}

As noted above, our temperature map is substantially in agreement with the
coarser map derived by Henry \& Briel (1996) using \rosat\ PSPC. The
\rosat\ map covers greater area than the \chandra\ data and shows
a hot sector extending to large radii in the NW. If this is correct, then a
comparison of the X-ray structure with some hydrodynamic simulations (e.g.\
Roetigger, Loken \& Burns 1997, hereafter RLB97) suggests that A2142 is the
result of an unequal merger, viewed at a time at least 1--2~Gyr after the
initial core crossing. The late phase is required by the largely smooth and
symmetrical structure of the X-ray emission, and the lack of obvious shocks.
In the simulations, shock-heated gas at the location of the initial impact
of the smaller system can still be seen at late times, similar to the hot
sector seen in the Henry \& Briel map far to the NW.  Hence in this model,
the low mass system has impacted from the NW.

The undisrupted cool core which we see in A2142 differs from what is seen in
the work of RLB97 and many others. However these simulations all involved
clusters with low core gas densities ($n<10^{-3}$~cm$^{-3}$). Under these
circumstances, the shock runs straight through the core of the main cluster,
raising its temperature. In contrast, it appears that the collision shock
has failed to penetrate the core of A2142, in which gas densities reach
$\sim 10^{-2}$~cm$^{-3}$, and has instead propagated around the outside,
heating the gas to the north and southwest of the cluster core.

In this model, galaxy G1 is identified with the center of the main cluster
(whose core includes the whole elliptical central region of A2142), and
there is less difficulty in accepting G2 (which lies essentially along the
collision axis, at least in projection) as being the former central galaxy
of the smaller subcluster. Having lost its gas halo on entering the cluster
from the NW, G2 has already crossed the center of the main cluster twice,
and is now either returning to the NW, or falling back towards the center
for a third time. The latter option derives some support from the fact that
the radio galaxy, presumably (from its similar line-of-sight velocity)
accompanying G2, has a narrow-angle radio tail which points to the west,
away from the center of the main cluster (Bliton et al.\ 1998).  The idea
that G2 has already crossed the cluster core also helps to explain the
elongated morphology of the central cooling flow, apparent in Fig.\ 3.

Simulations show that as the subcluster recrosses the cluster core, gas
which has been pulled out to the SE should fall in behind it, forming an
extended inflowing plume (see, e.g. RLB97 Fig.\ 8{\em f}). This is
consistent with the shallow X-ray surface brightness gradient seen to the SE
in Fig.\ 2{\em b}. In the case of A2142, this gas, flowing in from the SE,
will run into the dense cool core surrounding G1 at subsonic velocity, and
this could give rise to the SE density step through a physical mechanism
similar to that discussed in the previous section, involving gas shear and
stripping at the interface.

In this scenario, the NW edge may be the fossilized remains of the initial
subcluster impact that took place here. Shock heating from this impact has
raised the entropy of the gas outside the core to the NW. The shock has
propagated into the core until the radius where the pressure in the core
matched the pressure driving the shock. Subsequently, the flow of the
shocked gas towards the SE has swept away the outer layer of the core where
the shock decayed, leaving the high entropy shocked gas in direct contact
with the low entropy unshocked core. Once the gas returns to a hydrostatic
configuration, this entropy step manifests itself as a jump in temperature
and density of the form seen, while the gas pressure would be continuous
across the edge. In contrast to the model from the previous section, little
relative motion of the gas to either side of the NW edge is expected at this
late merger stage, so there is no current stripping. Simulations are
required to investigate how long a sharp edge of the kind observed can
persist under these conditions; this depends in part on poorly understood
factors such as the thermal conductivity of the gas.

\section{SUMMARY}

We have presented the results of a short \chandra\ observation of the
merging cluster A2142, which include a temperature map of its central region
and the temperature and density profiles across the two remarkable surface
brightness edges. The data indicate that these edges cannot be shock fronts
--- the dense gas inside the edges is cooler than the gas outside. It is
likely that the edges delineate the dense subcluster core(s) that survived
merger and shock heating of their surrounding, less dense atmospheres. We
propose that the edges themselves are surfaces where these cores are
presently being ram pressure-stripped by the surrounding hot gas, or
fossilized remains of such stripping which took place earlier in the merger.
More accurate temperature and pressure profiles for the edge regions would
help to determine whether the gas stripping is continuing at present, and
may also provide information on the gas thermal conductivity. A
comprehensive galaxy velocity survey of the cluster, and large-scale
temperature maps such as will be available from \xmm, will help to construct
a definitive model for this interesting system.

An accurate quantitative interpretation of the available optical and X-ray
data on A2142 requires hydrodynamic simulations of the merger of clusters
with realistically dense cores and radiative cooling. We also hope that the
results presented here will encourage an improvement in linear resolution of
the simulations necessary for modeling the sharp cluster features such as
those \chandra\ can now reveal.

\acknowledgements

The results presented here are made possible by the successful effort of the
entire \chandra\ team to build, launch and operate the observatory. Support
for this study was provided by NASA contract NAS8-39073 and by Smithsonian
Institution. TJP, PEJN and PM thank CfA for hospitality during the course of
this study.

\begin{figure*}[b]
\pspicture(0,2)(9,12)

\rput[tl]{0}(-0.1,12.3){\epsfxsize=9cm
\epsffile{lc.ps_b}}

\rput[tl]{0}(0,3.1){
\begin{minipage}{8.75cm}
\small\parindent=3.5mm
{\sc Fig.}~1.---Light curve for a region of chip S3 far from the cluster
brightness peak. Bins are 130 s. Shaded intervals of high (or low)
background are excluded from the analysis.
\par
\end{minipage}
}
\endpspicture
\end{figure*}

\begin{figure*}[b]
\pspicture(0,12)(18.5,24)

\rput[tl]{0}(0,24.1){\epsfysize=9.5cm \epsfclipon
\epsffile{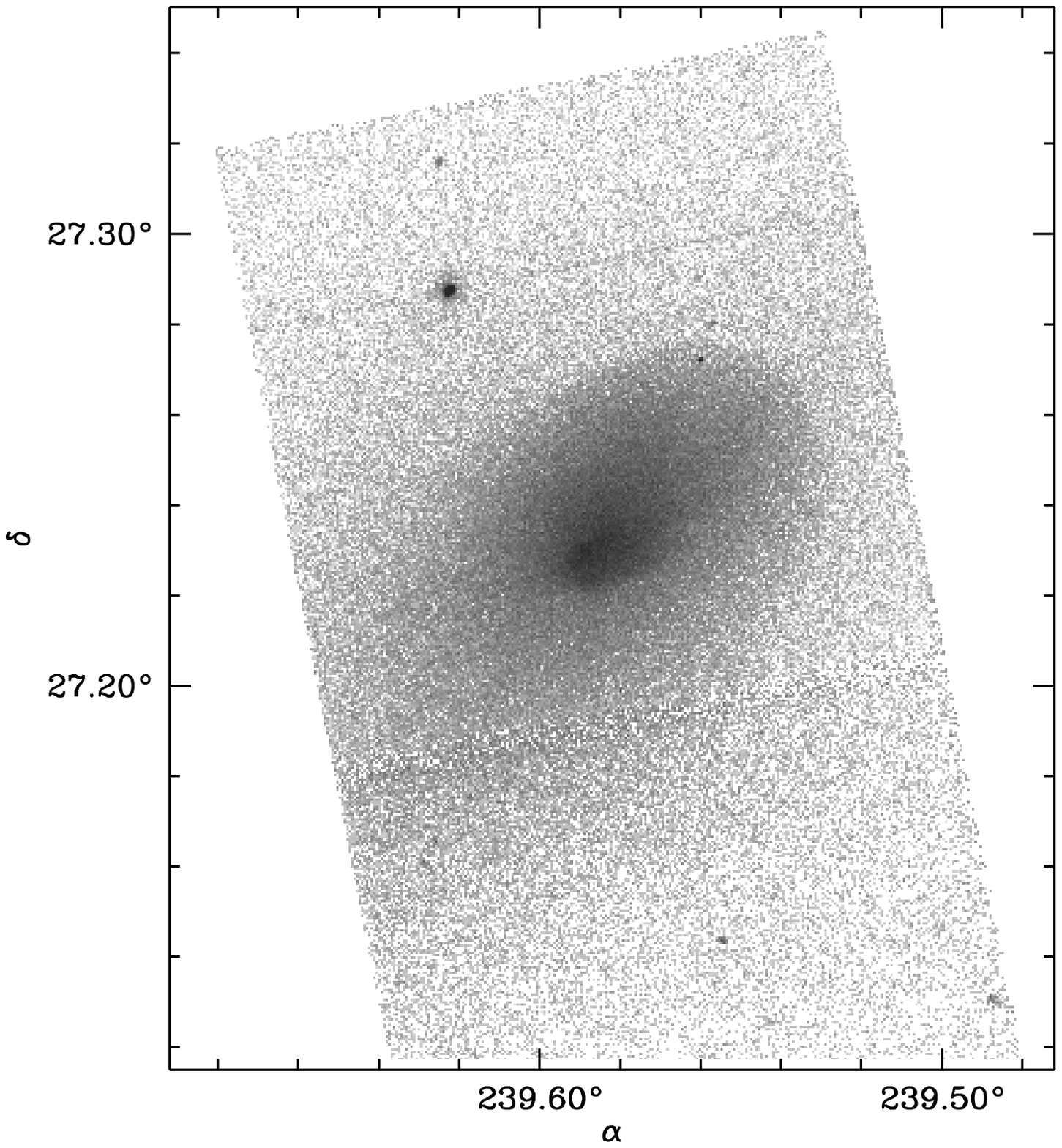}}

\rput[tl]{0}(8.5,24){\epsfysize=9.75cm
\epsffile{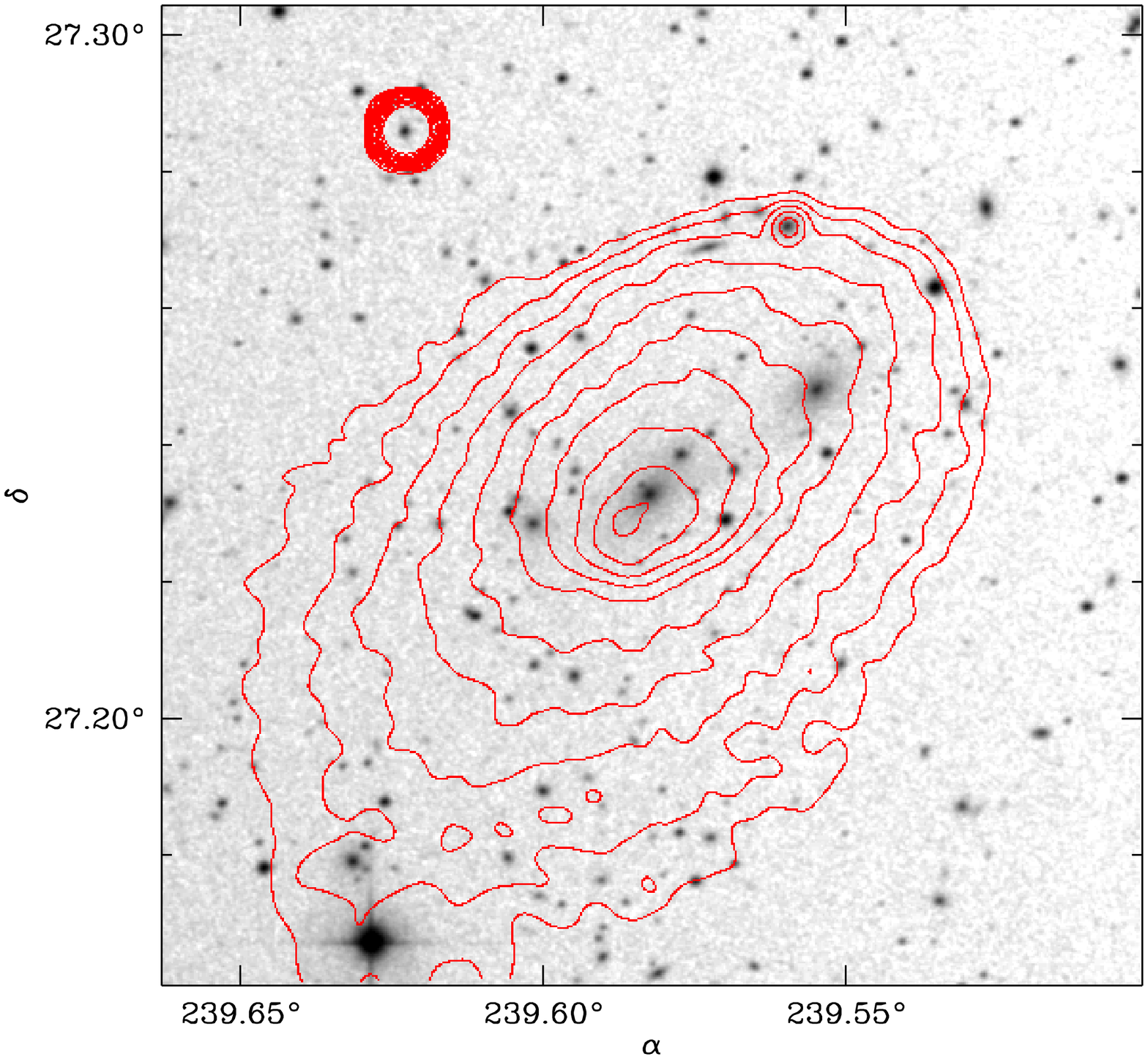}}

\rput[bl]{0}(2.0,23){\large\bf\it a}
\rput[bl]{0}(10.5,23){\large\bf\it b}

\rput[tl]{0}(0,14.0){
\begin{minipage}{18.5cm}
\small\parindent=3.5mm
{\sc Fig.}~2.---({\em a}) ACIS image of A2142 in the 0.3--10 keV band,
binned to 2\as\ pixels and divided by the vignetting map. Only chips S2 and
S3 are included. Note the two sharp elliptical brightness edges northwest
and south of the cluster peak. A streak of the emission that goes through
the bright point source (a Seyfert galaxy -- member of the cluster) is the
uncorrected track of this source during the CCD frame transfer. ({\em b}) A
Digitized Sky Survey image with overlaid ACIS X-ray brightness contours
(log-spaced by a factor of $\sqrt 2$). The two major galaxies, G1 and G2,
are seen at $\alpha=$239.\degd 5836, $\delta=$27.\degd 2333 and
$\alpha=$239.\degd 5556, $\delta=$27.\degd 2479, respectively.
\par
\end{minipage}
}
\endpspicture
\end{figure*}

\begin{figure*}[b]
\pspicture(0,2)(9,12)

\rput[tl]{0}(-0.2,12.3){\epsfxsize=9.6cm
\epsffile{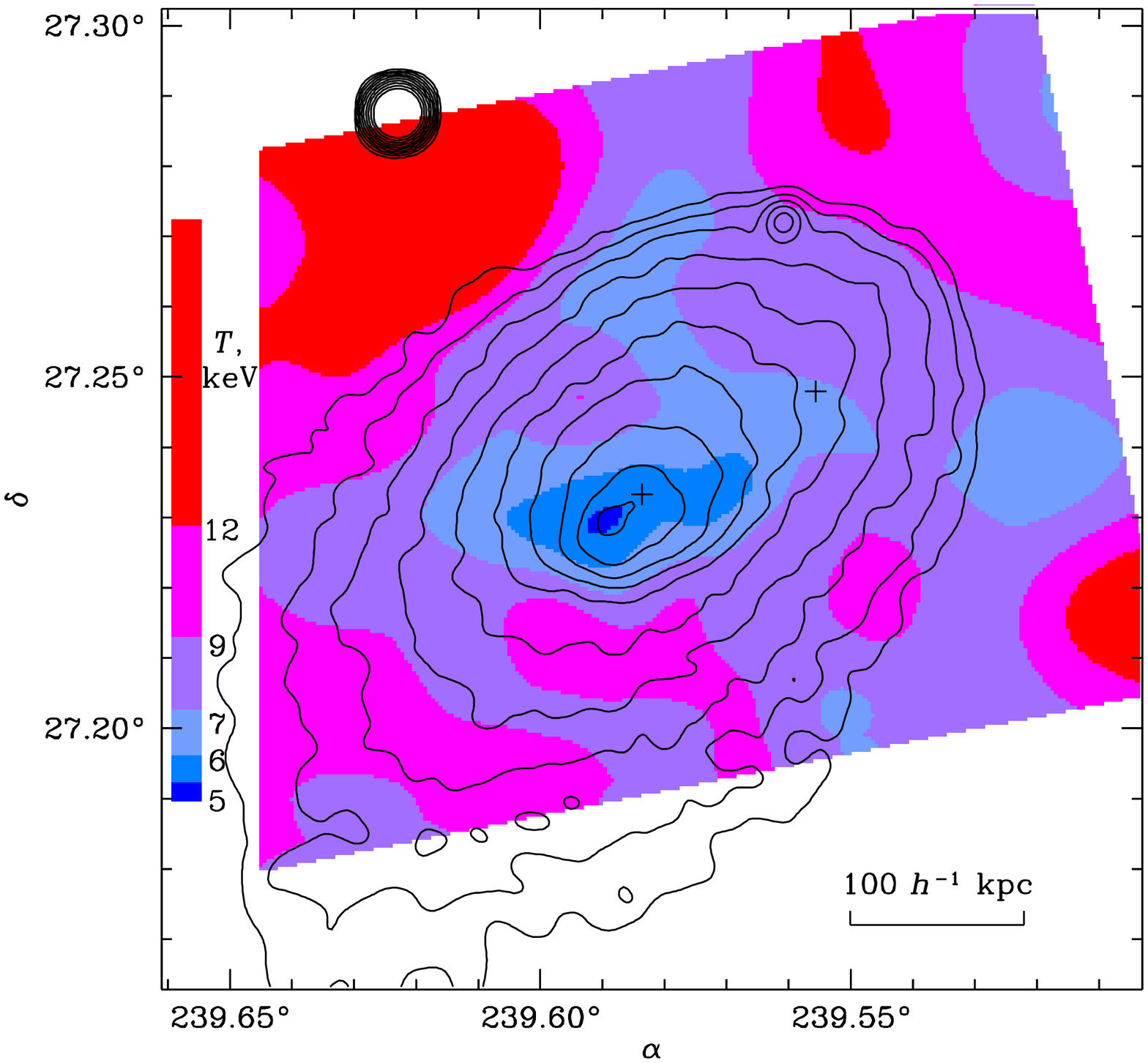}}

\rput[tl]{0}(0,3.7){
\begin{minipage}{8.75cm}
\small\parindent=3.5mm
{\sc Fig.}~3.---Temperature map of the central region of A2142 (color)
overlaid on the 0.3-10 keV ACIS brightness contours. The 90\% temperature
uncertainties increase from approximately $\pm0.5$ keV at the brightness
peak to $\pm1.5$ keV at the outer contour and still greater outside that
contour.  Crosses denote positions of the two brightest galaxies G1 (center)
and G2 (northwest).
\par
\end{minipage}
}
\endpspicture
\end{figure*}

\begin{figure*}[b]
\pspicture(0,2.5)(18.5,24)

\rput[tl]{0}(-0.1,24.0){\epsfxsize=9.5cm \epsfclipon
\epsffile{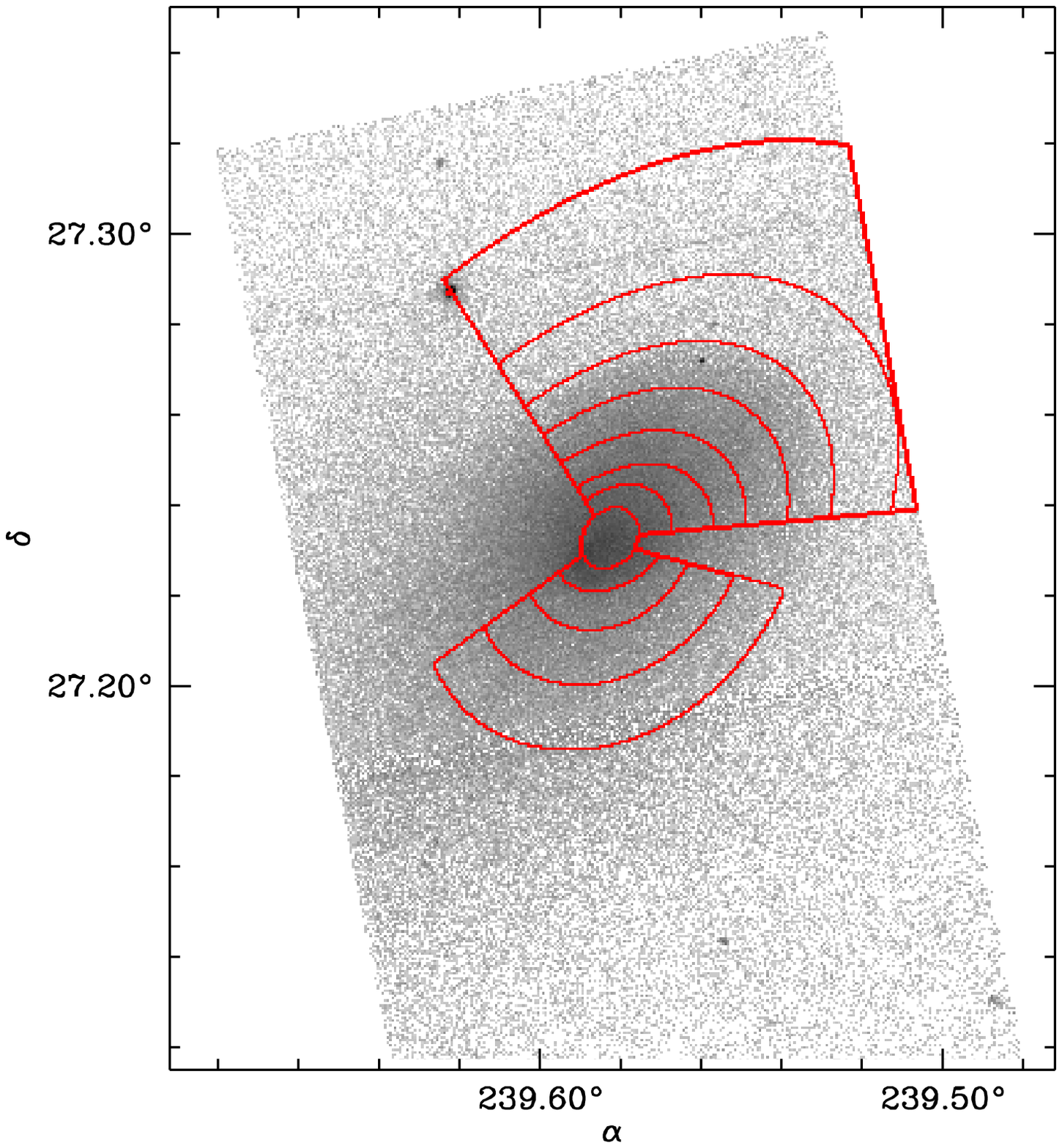}}

\rput[bl]{0}(-0.1,5.5){\epsfxsize=9cm
\epsffile{tprof.ps_b}}

\rput[tl]{0}(9.5,26.2){\epsfxsize=9cm
\epsffile{briprof.ps_b}}

\rput[tl]{0}(9.5,20.35){\epsfxsize=9cm
\epsffile{densprof.ps_b}}

\rput[bl]{0}(9.5,5.5){\epsfxsize=9cm
\epsffile{presprof.ps_b}}

\rput[bl]{0}(1.7,22.5){\Large\bf\it a}
\rput[bl]{0}(1.3,12.4){\Large\bf\it b}
\rput[bl]{0}(17.8,22.5){\Large\bf\it c}
\rput[bl]{0}(17.8,16.6){\Large\bf\it d}
\rput[bl]{0}(17.8,10.8){\Large\bf\it e}

\rput[tl]{0}(0,5.7){
\begin{minipage}{18.5cm}
\small\parindent=3.5mm
{\sc Fig.}~4.---({\em a}) Cluster X-ray image (same as in Fig.\ 2{\em a});
red overlay shows regions centered on the main galaxy G1 and used for
derivation of temperature profiles presented in panel ({\em b}). The
boundaries are chosen to highlight the brightness edges. Bright point
sources are excluded from the regions (not shown for clarity). In panels
({\em b-e}), the southern edge is shown in left plot and the northwestern
edge is in right plot (the central bin is same for both). In panel ({\em
b}), solid crosses show simultaneous temperature fits to both observations
and dashed crosses show separate fits to each observation. Errors are 90\%
and, for simultaneous fits, include background uncertainties. The
$r$-coordinate for the elliptical sectors corresponds to the
emission-weighted distance from the center. Panel ({\em c}) shows X-ray
brightness profiles across the edges. They are derived using sectors
parallel to the elliptical boundaries in panel ({\em a}) but with a finer
step (for the southern edge, we also used a somewhat narrower wedge angle
for sharpness). Data points are shown as 90\% error bars; the histogram is
the best-fit brightness model that corresponds to the gas density model
shown in panel ({\em d}). Panel ({\em e}) shows pressure profiles obtained
from the temperature and density profiles from panels ({\em b}) and ({\em
d}). Vertical dashed lines show the best-fit positions of the density jumps.
\par
\end{minipage}
}
\endpspicture
\end{figure*}

\begin{figure*}[b]
\pspicture(0,0)(9,12)

\rput[tl]{0}(-0.4,12){\epsfxsize=10cm
\epsffile{scheme.ps_b}}

\rput[cc]{0}(0.5,9.8) {\large\it a}
\rput[cc]{0}(0.5,4.6){\large\it b}

\rput[tl]{0}(0,1.4){
\begin{minipage}{8.75cm}
\small\parindent=3.5mm
{\sc Fig.}~5.---A model for A2142 proposed in \S\ref{sec:d1} is shown
schematically in panel ({\em b}). The preceding stage of the merger is shown
in panel ({\em a}). In panel ({\em a}), shaded circles depict dense cores of
the two colliding subclusters (of course, in reality, there is a continuous
density gradient). Shock fronts 1 and 2 in the central region of panel ({\em
a}) have propagated to the cluster outskirts in panel {\em b}, failing to
penetrate the dense cores that continue to move through the shocked gas
(core B is likely to be projected onto, or only grazing, core A). The cores
may develop additional shock fronts ahead of them, shown by dashed lines.
Scales and angles are arbitrary.
\par
\end{minipage}
}
\endpspicture
\end{figure*}

\end{document}